\documentclass[twocolumn,aps,pre,floatfix]{revtex4} 
\usepackage{graphicx}
\usepackage{color}
\usepackage{dcolumn}
\usepackage{bm}
\usepackage{amsmath}
\usepackage{soul}

\newcommand\beq{\begin{equation}}
\newcommand\eeq{\end{equation}}
\newcommand\bear{\begin{eqnarray}}
\newcommand\eear{\end{eqnarray}}

\begin{document}
\title{Variable-cell method for stress-controlled jamming of athermal, frictionless grains}
\author{Kyle C. Smith}
\email{kyle.c.smith@gmail.com}
\affiliation{Department of Materials Science and Engineering, Massachusetts Institute of Technology, Cambridge, Massachusetts 02139, USA}
\author{Ishan Srivastava}
\author{Timothy S. Fisher}
\email{tsfisher@purdue.edu}
\affiliation{Birck Nanotechnology Center and School of Mechanical Engineering,\\Purdue University, West Lafayette, Indiana 47907, USA}
\author{Meheboob Alam}
\affiliation{Engineering Mechanics Unit,\\Jawaharlal Nehru Centre for Advanced Scientific Research, Jakkur P.O., Bangalore 560064, India}
\begin{abstract}
{A new method is introduced to simulate jamming of polyhedral grains under controlled stress that incorporates global degrees of freedom through the metric tensor of a periodic cell containing grains.  Jamming under hydrostatic/isotropic stress and athermal conditions leads to a precise definition of the ideal jamming point at zero shear stress.  The structures of tetrahedra jammed hydrostatically exhibit less translational order and lower jamming-point density than previously described `maximally random jammed' hard tetrahedra.  Under the same conditions, cubes jam with negligible nematic order.  Grains with octahedral symmetry jam in the large-system limit with an abundance of face-face contacts in the absence of nematic order.  For sufficiently large face-face contact number, percolating clusters form that span the entire simulation box.  The response of hydrostatically jammed tetrahedra and cubes to shear-stress perturbation is also demonstrated with the variable-cell method.}
\end{abstract}
\maketitle


\section{Background}

The distinction among the various degrees of jamming has been made by Torquato and co-workers for hard grains \cite{TorJPCB2001}, and the mapping of such conditions to soft particle systems has been suggested \cite{HerPRE2003,DonPRE2004,HerPRE2004}.  In collectively jammed systems no collective motions can occur without overlapping other grains, while strictly jammed systems prohibit non-overlapping global motions (e.g., shear and cell extension) in addition to the collective condition.  In contrast, soft-grain jammed systems have been viewed from a mechanical paradigm in which the influence of global deformations has not been explored widely.  Approaches for simulating the athermal (i.e., zero temperature) jamming of soft spheres were pioneered by O'Hern and co-workers \cite{HerPRE2003} by employing structural optimization and molecular dynamics with fixed-cell shape.  Jamming has recently been simulated with similar methods for systems of various grain shapes, including ellipses \cite{MailPRL2009}, Platonic solids \cite{SmiPRE2010,SmiPRE2011,SmiJHT2012}, arbitrarily-shaped polyhedral grains \cite{SmiPCCP2012}, and irregular, polydispersed grains \cite{SmiIJHE2012}.  While soft and hard systems of spheres have exhibited consistency among their jammed densities \cite{HerPRE2003,DonPRE2004,HerPRE2004}, disparities have been observed between simulated soft and hard-tetrahedra packings \cite{SmiPRE2010,SmiJHT2012,JiaPRE2011} and those of experiments \cite{NeuAR2012,JaoPRL2010}.  Attempts have also been made to correlate the fraction of `face-jointed' (a stricter condition than a face-face contact, as in Ref.~\onlinecite{SmiPRE2011}) tetrahedra to density \cite{LiSM2013}.  The extrapolated minimal density (0.625 \cite{LiSM2013}) is very similiar to the jamming threshold density of athermal, soft tetrahedra (0.62-0.64 \cite{SmiPRE2010,SmiPRE2011}). 

Aside from the hard/soft nature of grains, the constraints imposed on granular systems during jamming (e.g., cell shape, stress, and temperature) are not consistent in the literature.  Only recently has the importance of global degrees of freedom in the jamming of two-dimensional disc packings been demonstrated \cite{DagPRL2012}, and methods are lacking for extension to three dimensions and arbitrary grain shapes.  In particular, Dagois-Bohy et al. \cite{DagPRL2012} recently demonstrated that jammed packing of disks, with unrelaxed global strain degrees of freedom, were unstable to shear with a negative shear modulus.  In addition, recent experiments suggest that metastable jammed states can form under shear \cite{BiNAT2011}.  Shape modulation of the periodic cell was an essential part of the densification protocol in hard-tetrahedra systems \cite{JiaPRE2011}, while cell shape was fixed in previous investigations of soft tetrahedra \cite{SmiPRE2010,SmiPRE2011}.  

Other simulations have probed the inherent structures determined by relaxation of randomly oriented soft grains in an isochoric, cube-shaped cell \cite{HerPRE2003,GaoPRE2006,XuSM2010}.  Such a process replicates the quenching of grains from a high-temperature state \cite{HerPRE2003}.  A fixed periodic-cell shape does not reflect the global degrees of freedom present in granular media, because this condition generally leads to non-zero shear stress.  In the large-system limit fixed-shape cells have been suggested to have negligible influence on the resultant jammed structures \cite{HerPRE2004}.  Also, O'Hern and co-workers have considered the effect of shear strain on jamming \cite{MailPRL2009,SchECT2010}, but shear deformations are restricted to specific planes (e.g., the $yz$, $xz$, and $xy$ planes of a cube-shaped periodic cell) within their shear-periodic framework (see Ref.~\onlinecite{LeeJPC1972}).  In contrast, variable cells parameterized in terms of lattice vectors, e.g., in Refs.~\cite{ParJAP1981,SouPRB1997,TorqJiaoPRE2009}, enable arbitrary shape deformations that are inherent to the definition of strict jamming \cite{TorJPCB2001}.    Based on such an approach, Jiao and Torquato \cite{JiaPRE2011} asserted that cell-shape variations account for the increased jamming threshold density (0.763) and face-face contact number ($>2$ per grain) relative to our previous results on systems jammed in fixed-shape cells (0.62-0.64 jamming threshold \cite{SmiPRE2010,SmiPRE2011} and $~1$ face-face contacts per grain \cite{SmiPRE2011}).

In contrast to granular systems, methods for the simulation of atomistic systems with stress as the controlled quantity have been developed by including cell volume \cite{AndJCP1980} and shape \cite{ParJAP1981,SouPRB1997} variations into system dynamics.  At equilibrium in a stress-controlled framework, enthalpy is minimized, and, when stress is hydrostatic, the framework yields an isobaric, isenthalpic ensemble \cite{ParJAP1981,SouPRB1997}.  Such a stress constraint at zero temperature replicates the $NPH$ ensemble \cite{ParJAP1981} commonly employed in molecular dynamics simulation.  In three dimensions, variable-cell deformations have been accommodated in simulations with a parallelepiped (i.e., triclinic) periodic cell whose shape is parameterized by a metric tensor \cite{SouPRB1997}.

In this article, a variable-cell method is introduced for simulating the athermal jamming of soft, frictionless grains under hydrostatic/isotropic loading.  In Sec.~\ref{sec:theory} a stress-controlled method is introduced for jamming of granular systems under arbitrary states of stress.   In Sections~\ref{sec:tetrahedra}, \ref{sec:cubes}, and \ref{sec:cuboct} the variable-cell jamming of tetrahedra, cubes, and grains with general octahedral symmetry is presented, respectively.  The ability of the new method to simulate jamming under both hydrostatic and shear loadings is demonstrated in Sections~\ref{sec:tetrahedra}.2 and \ref{sec:cubes}.

\section{Enthalpy-based variable-cell simulation method for jamming}\label{sec:theory}

\begin{figure}
\centering
\includegraphics[width=6.5cm]{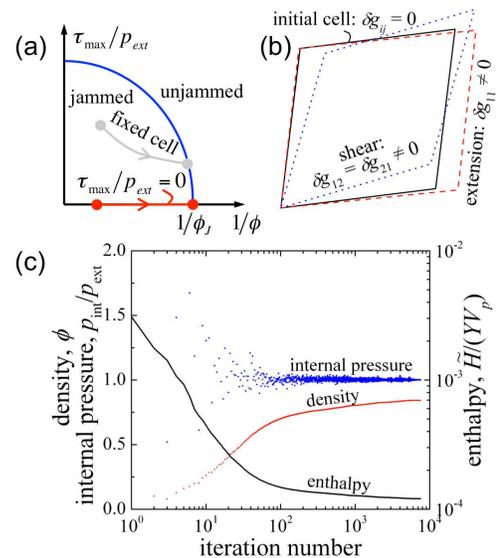}
\caption{(a) Granular phase-diagram \cite{LiuNAT1998,HerPRE2003} depicting a hydrostatic unloading process and a fixed cell-shape expansion process, both of which are at zero temperature.  (b) Irrotational global deformations are admitted through variations of the metric tensor describing the symmetric cell matrix $\tilde{\boldsymbol{h}}$.  Solid lines represent the edges of the initial cell that can undergo shear and extension to form the cells marked by dotted and dashed lines, respectively.  Each of these sample deformations correspond to changes in particular components of the metric tensor $\delta g_{ij}$. (c) Variation of density $\phi$, internal pressure $p_{int}$, and generalized enthalpy $\tilde{H}$ with global minimization iterations for a system of 1600 tetrahedra with an external pressure of $p_{ext}=10^{-4}Y$.}
\label{fig:PhsDia_Defor_Sam}
\end{figure}

Jamming, or the emergence of rigidity from a liquid state, may occur under conditions of non-zero shear stress, while the ideal \textit{jamming point} represents the liquid-solid transition in which temperature and shear stress vanish.  The density (i.e., volume fraction) at the \textit{jamming point} is the jamming threshold density $\phi_{J}$.   Because jamming processes in which the periodic-cell shape is fixed cannot control shear stress directly, such processes will generally follow a path that does not approach the \textit{jamming point} when expanded from a jammed state at finite pressure [Fig.~\ref{fig:PhsDia_Defor_Sam}(a), gray path].  Along such a path, the system will liquify at a density that differs from the jamming threshold density.  The principal stresses in a granular material must be identical with no shear-stress component (i.e., exactly hydrostatic or isotropic) to approach the \textit{jamming point} [Fig.~\ref{fig:PhsDia_Defor_Sam}(a), red path].

To simulate jamming under hydrostatic conditions (as well as, structural response subject to arbitrary applied stresses) the variable-cell method of Souza and Martins \cite{SouPRB1997} is employed here.   An arbitrary periodic-cell shape is defined by three lattice vectors, $\boldsymbol{a}_1$, $\boldsymbol{a}_2$, and $\boldsymbol{a}_3$, whose triple forms a 3-by-3 cell matrix $\boldsymbol{h}=[\boldsymbol{a}_1,\boldsymbol{a}_2,\boldsymbol{a}_3]$.  To perform structural optimization efficiently, cell-shape degrees of freedom are represented by the metric tensor: $\boldsymbol{g}=\boldsymbol{h}^T\boldsymbol{h}$ \cite{SouPRB1997}.  Because the metric tensor is symmetric, optimization needs only be performed over six of its nine elements.  Arbitrary shape variations are admissible through changes in components of the metric tensor $\delta g_{ij}$, including shear and extension [Fig.~\ref{fig:PhsDia_Defor_Sam}(b)].   Use of the metric tensor also prevents the occurence of artificial rigid-body rotations during the optimization routine \cite{SouPRB1997}.  Since non-spherical grains have rotational degrees of freedom, the mapping between metric tensor and cell matrix deformations should be irrotational.  Though an infinite set of cells, specified by $\boldsymbol{h}$, can satisfy such irrotational constraints, a unique and simple choice is the symmetric cell matrix $\tilde{\boldsymbol{h}}=\boldsymbol{g}^{1/2}$ \footnote{This can be evaluated with the diagonalized form of $\boldsymbol{g}$.}.  

The shape of the variable cell, given by $\boldsymbol{g}$, and the translational/rotational positions of all grains in the assembly, $\boldsymbol{r}$, evolve subject to the external state-of-stress $\boldsymbol{\sigma}_{ext}$ applied to the cell.  The Hamiltonian of this system includes the work done by $\boldsymbol{\sigma}_{ext}$, in addition to the internal energy $U(\boldsymbol{r})$ arising from elastic inter-grain contacts.  The quasi-static Hamiltonian at zero temperature (i.e., where kinetic energy is negligible) is the generalized enthalpy $\tilde{H}(\boldsymbol{r},g_{ij})$ \cite{SouPRB1997}:

\begin{equation}
  \tilde{H}(\boldsymbol{r},g_{ij})=U(\boldsymbol{r})+p_{ext}\sqrt{\text{det}g_{ij}}+g_{ij}\tau_{ext}^{ij},
\end{equation}

\noindent where $p_{ext}$ and $\boldsymbol{\tau}_{ext}$ are the external pressure and deviatoric-stress tensor that define the external state-of-stress, $\boldsymbol{\sigma}_{ext}=p_{ext}\boldsymbol{I}_{3}+\boldsymbol{\tau}_{ext}$ ($\boldsymbol{I}_{3}$ is the $3\times3$ indentity matrix).  Superscript Einstein indices $kl$ of a given tensor $\boldsymbol{\sigma}$ denote a particular contravariant-lattice component $\sigma^{kl}$ of that tensor \footnote{The contraviant-lattice representation can be determined from a given tensor's Cartesian representation $\boldsymbol{\sigma}_{\text{cart}}$ as $[\sigma^{kl}]=(\text{det}h)h^{-1}\boldsymbol{\sigma}_{\text{cart}}(h^{-1})^T$ \cite{SouPRB1997}.}.

Gradients of the enthalpy with respect to metric coordinates $g_{ij}$ must also be expressed to optimize structure \cite{SouPRB1997}:

\begin{equation}
  \frac{\partial \tilde{H}}{\partial g_{ij}}=-\frac{1}{2}\sigma_{net}^{ij},
  \label{eq:dHdgij}
\end{equation}

\noindent where the net stress tensor $\boldsymbol{\sigma}_{net}$ is simply the difference between the internal $\boldsymbol{\sigma}_{int}$ and external $\boldsymbol{\sigma}_{ext}$ stresses:

\begin{equation}
  \boldsymbol{\sigma}_{net}=\boldsymbol{\sigma}_{int}-\boldsymbol{\sigma}_{ext}.
\end{equation}

\noindent  The internal stress tensor $\boldsymbol{\sigma}_{int}$ is a sum over each contact $k$:

\begin{equation}
  \boldsymbol{\sigma}_{int} = \frac{1}{\sqrt{\text{det}g_{ij}}}\sum_{k}{\boldsymbol{F}_{k}\otimes\boldsymbol{l}_{k}},
\end{equation}

\noindent where $\boldsymbol{F}_{k}$ and $\boldsymbol{l}_{k}$ are the force and branch vector of contact $k$, respectively.  In systems of non-spherical grains $\boldsymbol{\sigma}_{int}$ is only symmetric when contact moments are in equilibrium \cite{BarIJSS2001}, but its anti-symmetric component vanishes as internal degrees of freedom equilibrate during enthalpy minimization.

Equivalent off-diagonal components of the metric tensor evolve identically during optimization (i.e., $\delta g_{ij}=\delta g_{ji}$), and, therefore, the contribution of both enthalpy gradients (given by Eq.~\ref{eq:dHdgij}) must be accounted for when optimizing the cell's shape.  The translational degrees of freedom of individual grains are optimized in terms of their lattice coordinates, instead of their Cartesian coordinates.  The lattice translational coordinates $\boldsymbol{s}_t$ are expressed in terms of their Cartesian coordinates $\boldsymbol{r}_t$ and the cell transformation matrix $\boldsymbol{h}$ as $\boldsymbol{s}_t=\boldsymbol{h}^{-1}\boldsymbol{r}_t$.  The enthalpy gradient with respect to the lattice coordinates of a given grain is $-\boldsymbol{h}^T\boldsymbol{F}$, where $\boldsymbol{F}$ is the net force on that grain.  Enthalpy gradients with respect to rotational degrees of freedom are equivalent to the internal energy gradients presented in Ref.~\cite{SmiPRE2010}.  Soft, repulsive pair potentials are employed to describe the conservative elastic interactions between grains.  For polyhedral grains an overlap potential $E_{\alpha\beta}$ is employed here that depends on the intersecting volume between contacting grains, $E_{\alpha\beta}=0.25Y V_{\alpha\beta}^{2}/V_p$ \cite{SmiPRE2010}, where $V_{\alpha\beta}$ is the intersecting volume between grains $\alpha$ and $\beta$, $V_p$ is the volume of an individual grain, and $Y$ is the elastic modulus of the grains.

Finally, the optimization of enthalpy generally proceeds in a computationally inefficient manner when only gradients are employed to inform iterative search directions (e.g., with the conjugate gradient method), because of the drastically contrasting stiffnesses (and even unlike units thereof) among global and internal degrees of freedom.  Therefore, the quasi-Newton L-BFGS algorithm \cite{NocMP1989} was employed to optimize enthalpy with the aid of a specialized data structure \cite{CarLBFGS}.  Diagonal pre-conditioning was also employed to maximize convergence rates of the iterative sequence.

\section{Results and Discussion}

The results for variable-cell jamming of tetrahedra, cubes, and grains with octahedral symmetry under conditions of hydrostatic stress, specified by external pressure $p_{ext}=10^{-4}Y$, are presented in Sec.~\ref{sec:tetrahedra}, \ref{sec:cubes}, and \ref{sec:cuboct}, respectively.  The isobaric shear-response of jammed systems of tetrahedra and cubes are also discussed.  Each realization of a given system was initialized with random grain positions and orientations at a dilute density (i.e., volume fraction of grains) of $\phi=0.001$ \footnote{Because the present method is stress-controlled, the choice of initial density is insignificant.  Specifically, if a density is chosen for which any appreciable contact occurs in the initialized state, expansion of the variable cell will occur (instantaneously reducing the density of the granular system) because of the lack of equilibrium between internal and external pressure.}.  During structural optimization cell shape automatically contracts and grain positions/orientations evolve subject to the external pressure applied [Fig.~\ref{fig:PhsDia_Defor_Sam}(c)].  As enthalpy $\tilde{H}$ is minimized the system's density converges toward an equilibrium value [Fig.~\ref{fig:PhsDia_Defor_Sam}(c)].  Except at very dilute densities the external stress is nearly in equilibrium with the internal force network, as evidenced by the near unity ratio of internal-to-external pressure, $p_{int}/p_{ext}$ [Fig.~\ref{fig:PhsDia_Defor_Sam}(c)].  To estimate the jamming threshold density, external pressure was reduced and contact depth was extrapolated, as described in Ref.~\onlinecite{SmiPRE2010}.

\subsection{Jamming of tetrahedra and the difference between variable- and fixed-cell simulation}\label{sec:tetrahedra}

The packing of tetrahedra has received significant attention experimentally and computationally.   Specifically, athermal jammed structures of soft, frictionless tetrahedra within a fixed-shape periodic-cell have exhibited strong similarity with the radial distribution functions (RDFs) of experimental packings of tetrahedron-like grains (see \cite{SmiPRE2010,SmiJHT2012}).  Jiao and Torquato \cite{JiaPRE2011} reported a density of $0.763\pm0.005$ and $2.21\pm0.01$ face-face contacts per grain in `maximally random jammed' packings of hard tetrahedra.  In contrast, we reported previously a jamming threshold density of $0.634\pm0.011$ (similar to that of monodisperse spheres) and approximately $1$ face-face contact per grain of athermal tetrahedra jammed within a cell of fixed shape \cite{SmiPRE2011}.

\subsubsection{Hydrostatic, variable-cell jamming: Translational order frustration}

\begin{figure}
\centering
\includegraphics[width=6.5cm]{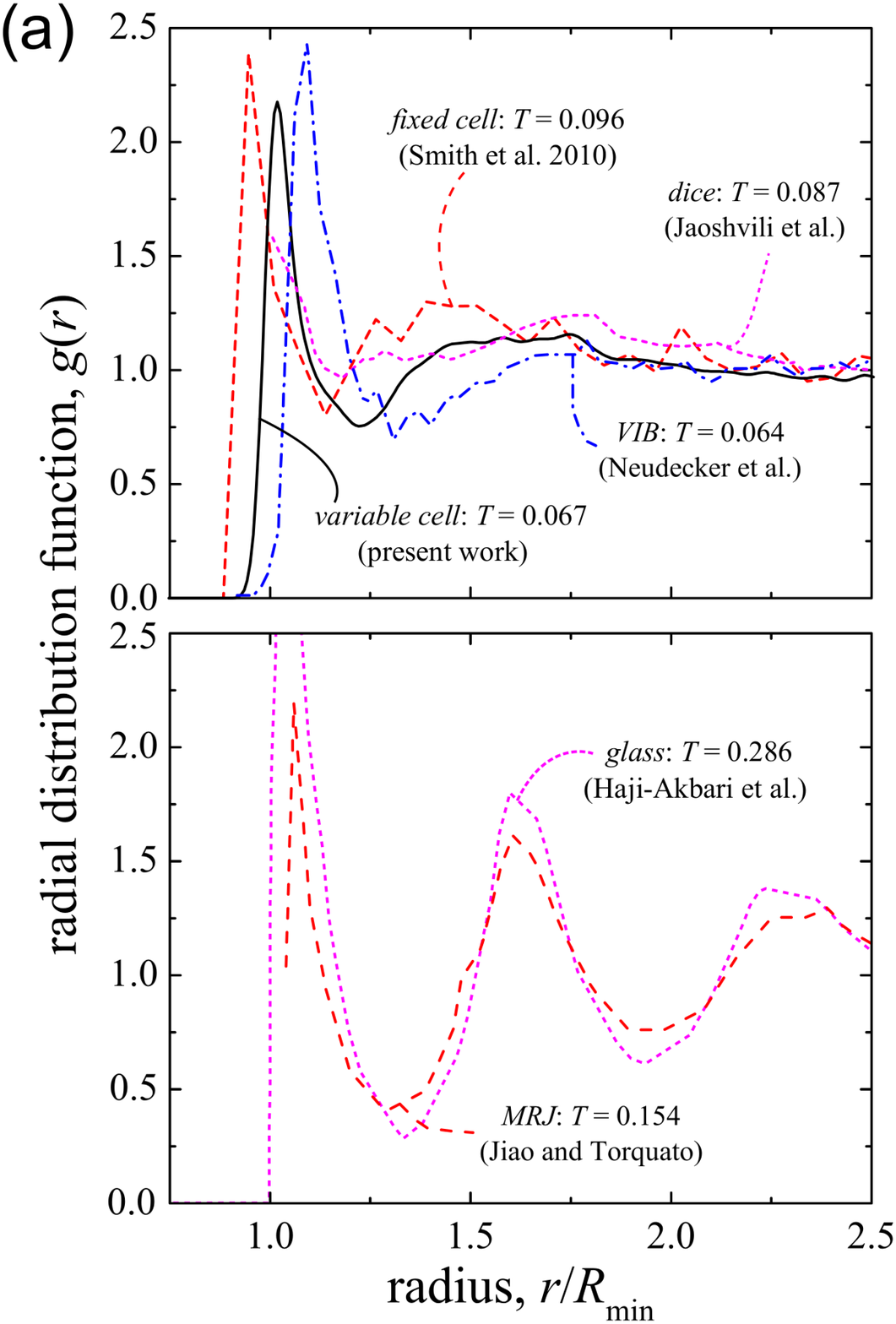}\\
\includegraphics[width=6.5cm]{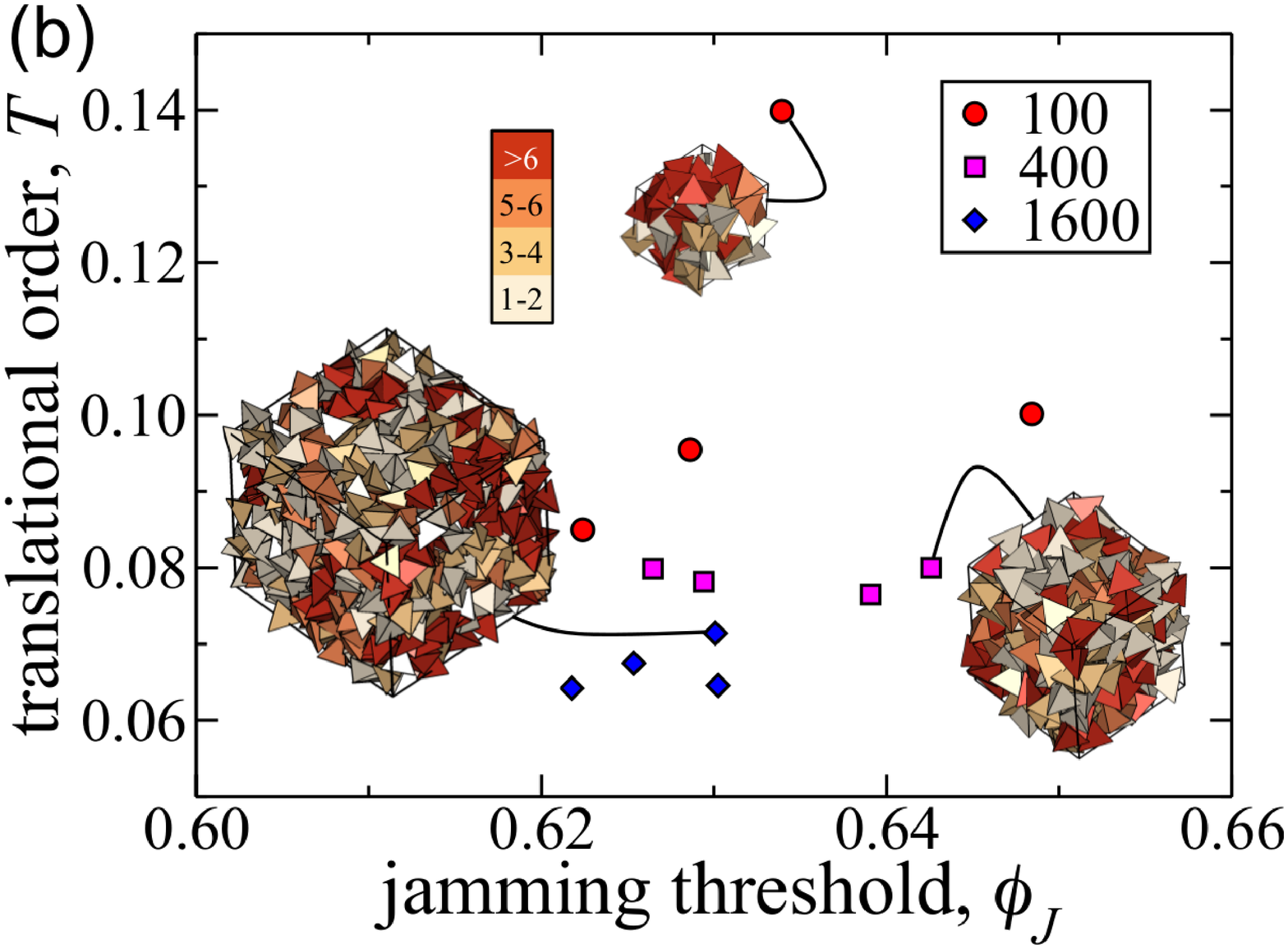}
\caption{(a) Radial distribution function (RDF) of hydrostatically jammed tetrahedra at $p_{ext}=10^{-4}Y$ and those of other simulated \cite{SmiPRE2010,AkbNat2009,JiaPRE2011} and experimental studies \cite{NeuAR2012,JaoPRL2010} from the literature.  $2R_{min}$ is the minimum separation between contacting tetrahedra.  The translational order parameter $T$, Eq.~\ref{Trans_ord_param}, is indicated for each system as well.  (b) Translational order $T$ as a function of jamming threshold $\phi_J$ for multiple realizations of variable cells of various sizes.  Selected realizations grains are colored according to the size of the face-face clusters to which they belong (see legend).  Black lines indicate periodic-cell boundaries.}
\label{fig:RDFs}
\end{figure}

To assess the randomness of jammed structures of tetrahedra, the RDF was computed and compared to previous simulated and experimental data [Fig.~\ref{fig:RDFs}(a)].   The present systems exhibit RDFs very similar to the experimental RDFs of vibrated tetrahedron-like grains \cite{NeuAR2012} and packed tetrahedral dice \cite{JaoPRL2010}, denoted respectively as \textit{VIB} and \textit{dice} in Fig.~\ref{fig:RDFs}.  Despite the initial peak that results from the steric replusion between tetrahedra, these RDFs exhibit only short-range order (i.e., the RDF is near unity beyond the initial peak).  In contrast, the RDFs of so-called `maximally random jammed' hard tetrahedra (\cite{JiaPRE2011}, denoted as \textit{MRJ}) and glassy tetrahedra (\cite{AkbNat2009}, denoted as \textit{glass}) exhibit several strong peaks absent in all other packings considered, indicating longer range order in the latter packings \cite{AkbNat2009,JiaPRE2011}.  The figure also includes the RDF from our prior fixed-cell athermal simulation of 400 tetrahedra \cite{SmiPRE2010}.  Comparison to the present variable-cell results confirms that the previous fixed-cell simulation predicts higher short-range order, as indicated by the peaks in the range $1\lesssim r/R_{min} \lesssim 2$.

The crystal-independent translational order parameter $T$ was also computed to quantify order in systems of tetrahedra \cite{TruPRE2000}:

\begin{equation}
\label{Trans_ord_param}
T = \frac{1}{r_{max}-r_{min}}\int_{r_{min}}^{r_{max}}|g(r)-1|dr.
\end{equation}

\noindent The values of $T$ shown in Fig.~\ref{fig:RDFs} were obtained by integrating the RDF from a particular $r_{min}$ \footnote{For the fixed-cell simulation from Ref.~\onlinecite{SmiPRE2010}, $r_{min}$ was chosen as the smallest $r$ for which $g(r)=1$, while for all other cases $r_{min}=R_{min}$.} up to $r_{max}=4R_{min}$ (where $2R_{min}$ is the minimum distance between non-overlapping tetrahedra).  Four realizations of each system size ($100$, $400$, and $1600$ tetrahedra) were simulated.  These results reveal that order generally decreases with system size [see Fig.~\ref{fig:RDFs}(b)].  On average the translational order parameter for systems of $1600$ tetrahedra is $0.064$, which is similar to those computed for the experimental structures also considered.  By extrapolation, we predict the infinite system to have $T=0.058$.  This order metric confirms that `maximally random jammed' hard tetrahedra (from Ref.~\onlinecite{JiaPRE2011}), as well as our prior results from fixed-cell jamming \cite{SmiPRE2010}, are, in fact, \textit{less random} than the present soft, frictionless tetrahedra jammed athermally in a variable cell under hydrostatic stress.  

Tetrahedra jammed in a variable cell exhibit an average jamming threshold density of $0.627$ [Fig.~\ref{fig:RDFs}(b)] -- similar to our previous observations on systems in cells of fixed, cubic shape.  This finding contrasts with the suggestion of Jiao and Torquato \cite{JiaPRE2011} that global degrees of freedom enable tetrahedra to jam with structures having higher jamming threshold density than our previous predictions \cite{SmiPRE2010,SmiPRE2011}.  The discrepancy between athermally jammed tetrahedra and hard tetrahedra \cite{JiaPRE2011} is likely related to the inherent thermalizing nature of the Monte Carlo-based method (Adaptive Shrinking Cell \cite{TorqJiaoPRE2009}) employed to sample the phase-space of hard-grain packings.  In a kinetic sense, thermalization enables packed systems to overcome barriers resulting from the low-density jammed structures formed in athermal systems.  This notion is also consistent with the granular phase-diagram for which higher jamming transition densities are expected for systems at finite temperature than for athermal systems \cite{LiuNAT1998,HerPRE2003}.

\subsubsection{Features that distinguish between structures jammed in variable and fixed cells: Jammed state-of-stress and shear response}

\begin{figure}
\centering
\includegraphics[width=6.5cm]{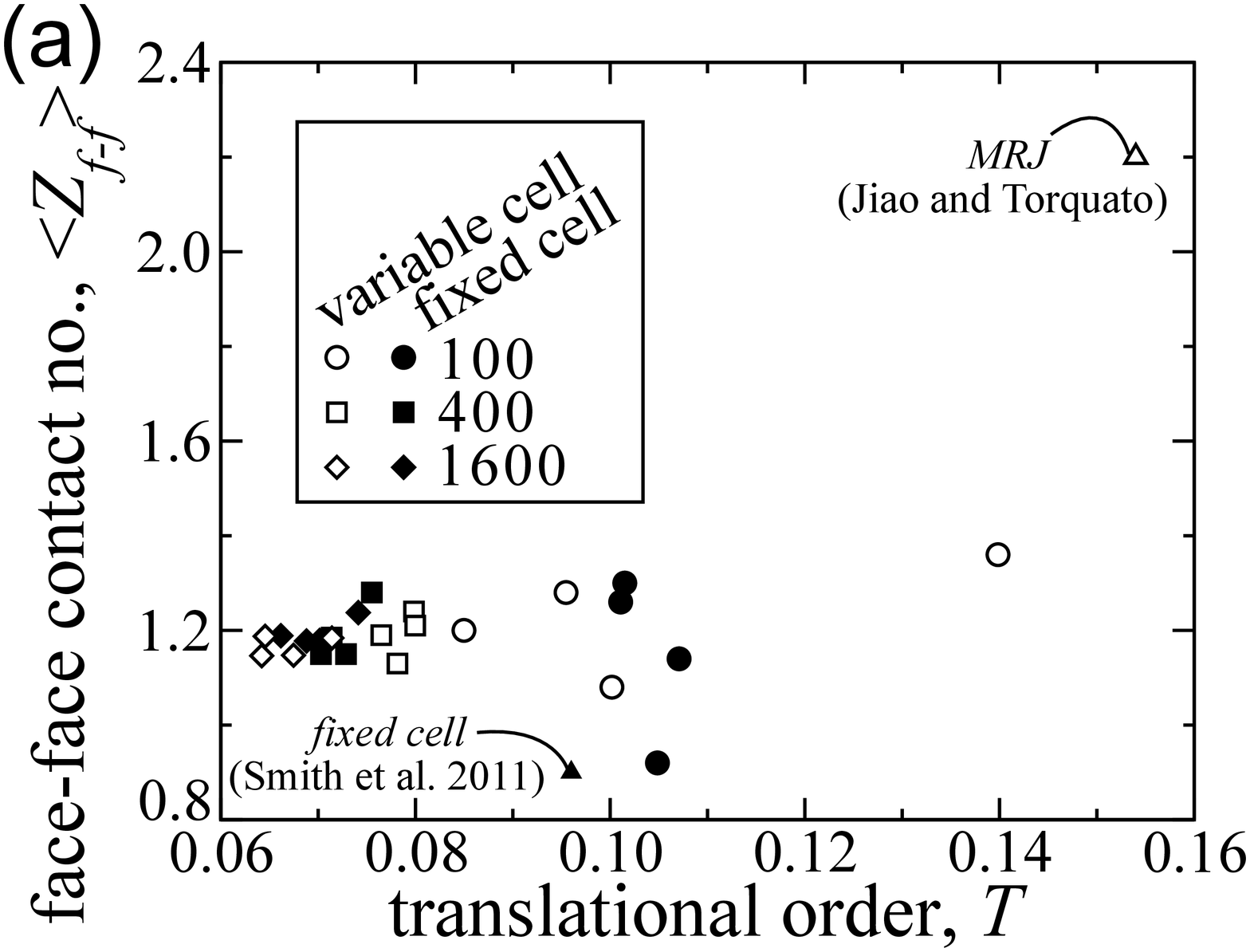}\\
\includegraphics[width=6.5cm]{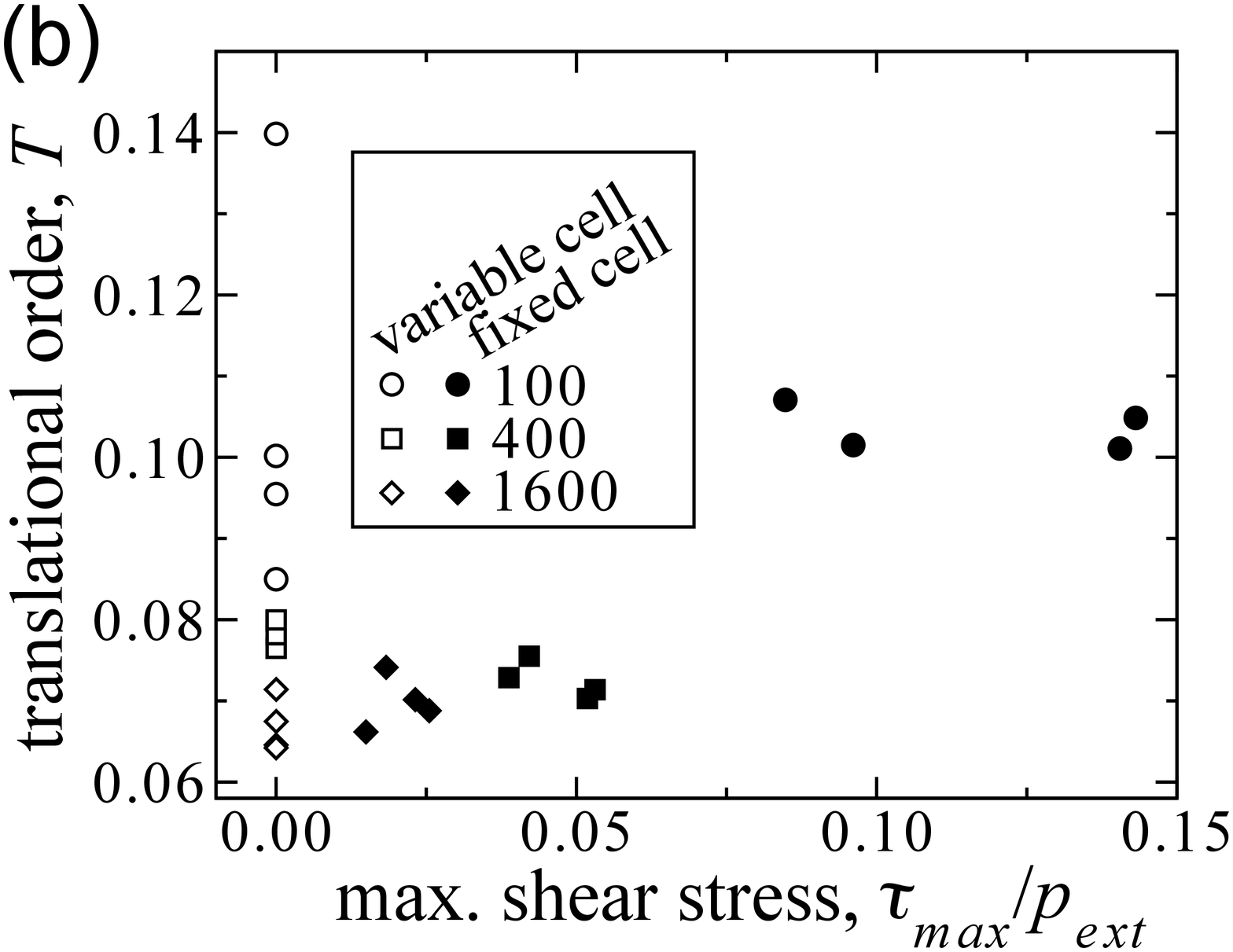}
\caption{Structural characteristics of configurations jammed in fixed and variable cells.  (a) Face-face contact number as a function of translational order $T$.  Previous results from Refs.~\onlinecite{SmiPRE2011,JiaPRE2011} are shown as reference points.  (b) Translational order parameter $T$ as a function of the maximum shear stress for systems jammed in variable and fixed cells.}
\label{fig:tetshear}
\end{figure}

Using the present stress-controlled approach, we have also generated jammed structures with  cubic-cell shape by performing constrained minimization of the generalized enthalpy (i.e., with $\boldsymbol{\sigma}_{ext}=p_{ext}\boldsymbol{I}_3$, $\delta g_{11}=\delta g_{22}=\delta g_{33} \neq 0$, and $\delta g_{23}=\delta g_{13}=\delta g_{12}=0$).  This approach for fixed-cell jamming differs from our previous approach (described in Ref.~\onlinecite{SmiPRE2010}) involving an alternating sequence of isochoric, internal-energy minimizations with affine, isotropic strains.  We first compare face-face contact numbers $\langle Z_{f-f} \rangle$ between tetrahedra jammed in fixed- and variable-shape cells (Fig.~\ref{fig:tetshear}(a), assuming that contacts with faces aligned by $<1^{\circ}$ are face-face contacts \cite{SmiPRE2011}).  We observe that fixed-cell shape results in variability of $\langle Z_{f-f} \rangle$ among small-system realizations, but in all cases (including both fixed- and variable-cell structures) $\langle Z_{f-f} \rangle$ is of similar magnitude to our previous predictions \cite{SmiPRE2011} but nearly half that reported for `maximally random jammed' tetrahedra \cite{JiaPRE2011}.  The grouping of $\langle Z_{f-f} \rangle\approx1.2$ demonstrates that face-face contact numbers of jammed tetrahedra are insensitive to the cell shape for sufficiently large systems (i.e., $\gtrsim400$ tetrahedra), contrary to previous claims by Jiao and Torquato \cite{JiaPRE2011}.

As shown in Fig.~\ref{fig:tetshear}(b), each realization (whether having a variable- or fixed-shape cell) has a corresponding translational order $T$ and maximum shear stress $\tau_{max}$ (i.e., half the difference between the maximum and minimum principal stresses of $\boldsymbol{\sigma}_{int}$).  The internal state-of-stress $\boldsymbol{\sigma}_{int}$ for fixed cells differs from the applied hydrostatic stress $p_{ext}\boldsymbol{I}_3$, because the fixed cell has only one global degree of freedom (the cell volume) instead of the six global degrees of freedom of the variable cell (i.e., $ g_{11}$, $ g_{22}$, $ g_{33}$, $ g_{23}$, $g_{13}$, and $g_{12}$).  $\tau_{max}$ depends on the particular configuration being jammed in a fixed cell and, consequently, exhibits statistical variation among multiple realizations of a given system size.  Despite this artificial and unpredictable effect on the mechanical state of the jammed system, the magnitude of $\tau_{max}$ decreases (on average) as the size of the fixed cell increases.  In other words, the state-of-stress in fixed cells approaches a hydrostatic state-of-stress for large systems.  As a result, the translational order predicted by fixed cells is very close to that of variable cells for sufficiently large systems, because they approach a common state-of-stress in this limit.

\begin{figure}
\centering
\includegraphics[width=6.5cm]{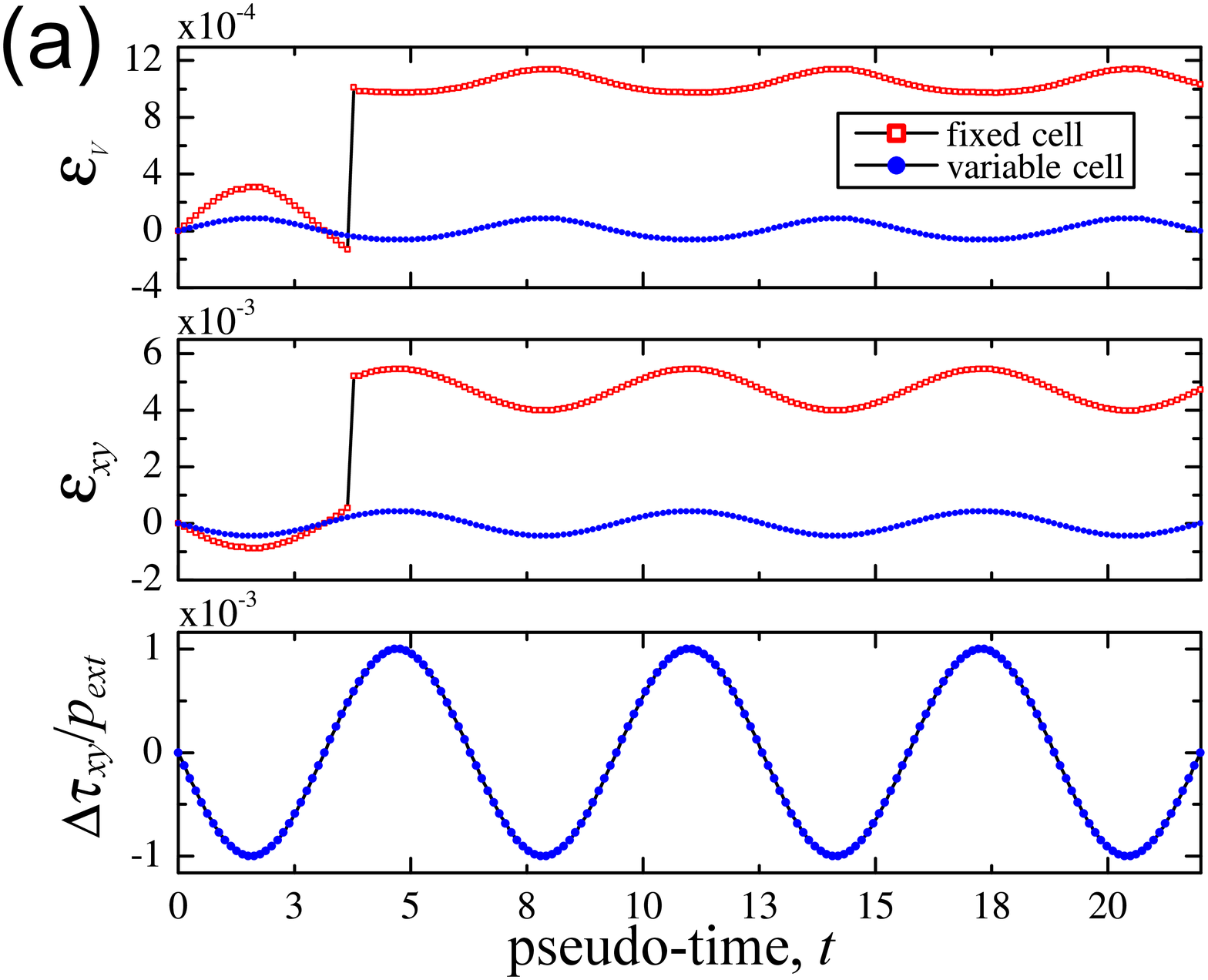}\\
\includegraphics[width=6.5cm]{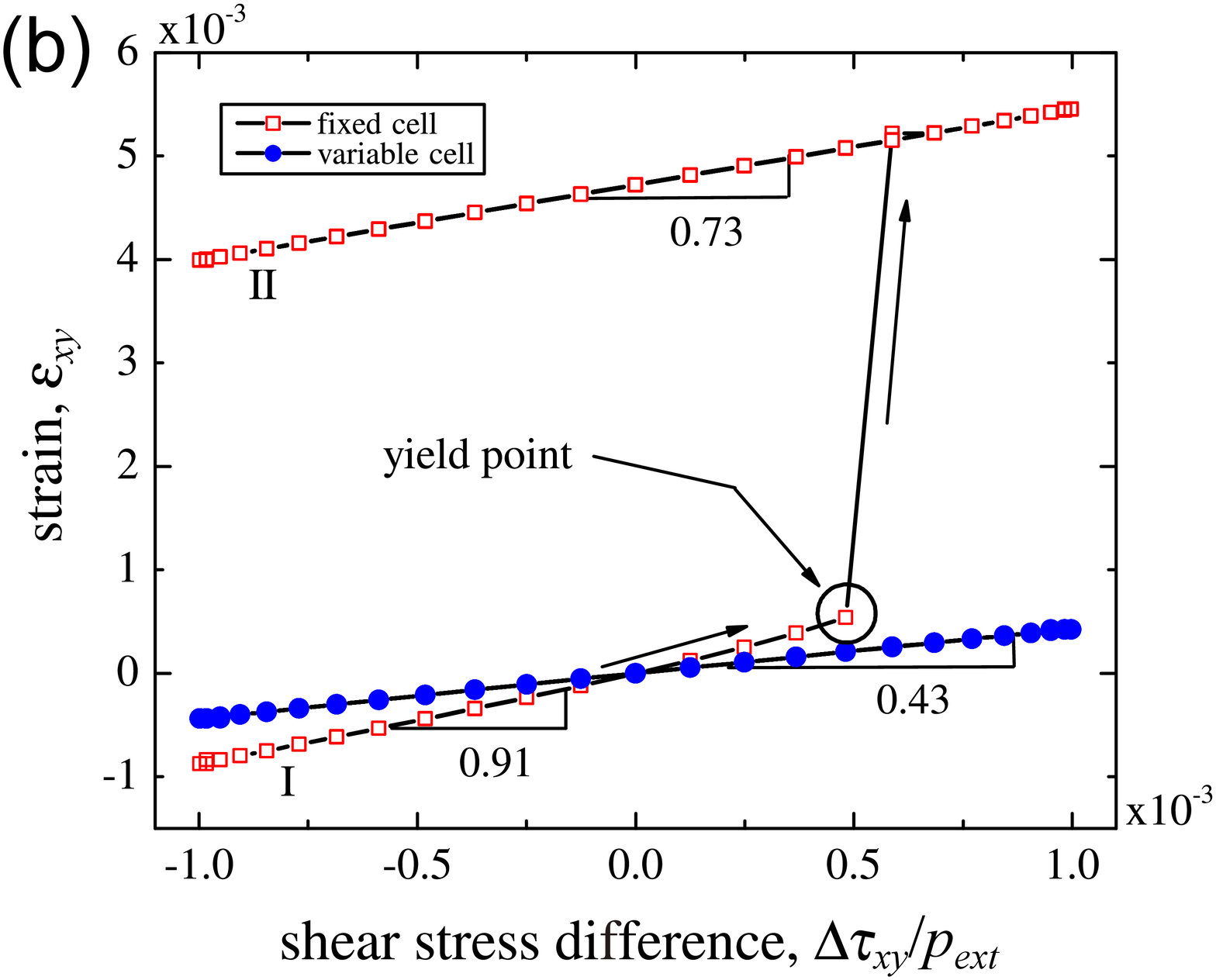}
\caption{Strain response to oscillatory shear stress of $100$ tetrahedra jammed in either a variable or fixed cell at $p_{ext}=10^{-4}Y$.  (a) Time evolution of shear strain $\epsilon_{xy}$ and volumetric strain $\epsilon_{V}$ with applied shear-stress difference $\Delta\tau_{xy}$.  (b) Lissajous-Bowditch curves of $\epsilon_{xy}$ versus $\Delta\tau_{xy}$.   For the fixed cell, the yield point is indicated (at which plastic flow ensues).}
\label{fig:tetshear2}
\end{figure}

The variable-cell method can also be used to simulate states of stress perturbed from the jammed system (i.e., that due to hydrostatic jamming in either a variable or fixed cell).  Using this capability, we applied an oscillatory shear-stress perturbtation and simulated the strain response with all global degrees of freedom active (i.e., such that $\boldsymbol{\sigma}_{int}=\boldsymbol{\sigma}_{ext}$ after enthalpy is minimized).  We refer to the stress-sequencing variable as `pseudo-time' $t$ in Fig.~\ref{fig:tetshear2}(a) because time is only a surrogate for the order in which stress perturbations occur for the present quasi-static simulations.   The $xy$ component of external stress was chosen to vary in psuedo-time as $\sigma_{ext,xy}(t)=\sigma_{ext,xy}^0+\Delta \tau_{xy}(t)$, where superscript $0$ denotes the jammed value of stress and $\Delta \tau_{xy}(t)$ is the shear-stress difference.  The remaining non-degenerate components of external stress were fixed at their jammed values at all instants in pseudo-time, i.e., $\sigma_{ext,ij}(t)=\sigma_{ext,ij}^0$.  The resulting strains were determined from the evolved metric tensor \footnote{The strain tensor was computed as $\boldsymbol{\epsilon}=0.5((\boldsymbol{h}_0^T)^{-1} \boldsymbol{g}\boldsymbol{h}_0^{-1}-\boldsymbol{I}_3)$ \cite{ParJAP1981}, where $\boldsymbol{h}_0$ is the hydrostatically jammed cell-matrix.}.

For the small-amplitude oscillations shown in Fig.~\ref{fig:tetshear2}(a), tetrahedra jammed in a variable cell respond elastically with negligible hysteresis.  In contrast, tetrahedra jammed in a fixed cell undergo yielding and an instantaneous plastic-flow event at $\Delta \tau_{xy}/p_{ext}\approx 5\times 10^{-4}$ [Fig.~\ref{fig:tetshear2}(a)].  This `yield point' separates two regimes of elasticity: pre- and post-yield (termed, I and II, respectively).  During the respective elastic regimes both systems undergo antiphase dilation with the applied shear-stress difference, but the plastic-flow event in the fixed cell contracts its volume [Fig.~\ref{fig:tetshear2}(a)].  Also, both the differential shear modulus $G=\partial\Delta\tau_{xy}/\partial\epsilon_{xy}$ and shear strength are enhanced by plastic flow for the fixed cell [Fig.~\ref{fig:tetshear2}(b)].  Additionlly, the differential shear-modulus of the variable cell exceeds that of the fixed cell in both regimes I and II [Fig.~\ref{fig:tetshear2}(b)].  These results demontrate that configurations jammed in fixed cells are weaker and more compliant than those jammed in variable cells.

\subsection{Jamming of cubes: Nematic order frustration and shear response}\label{sec:cubes}

\begin{figure}
\centering
\includegraphics[width=6.5cm]{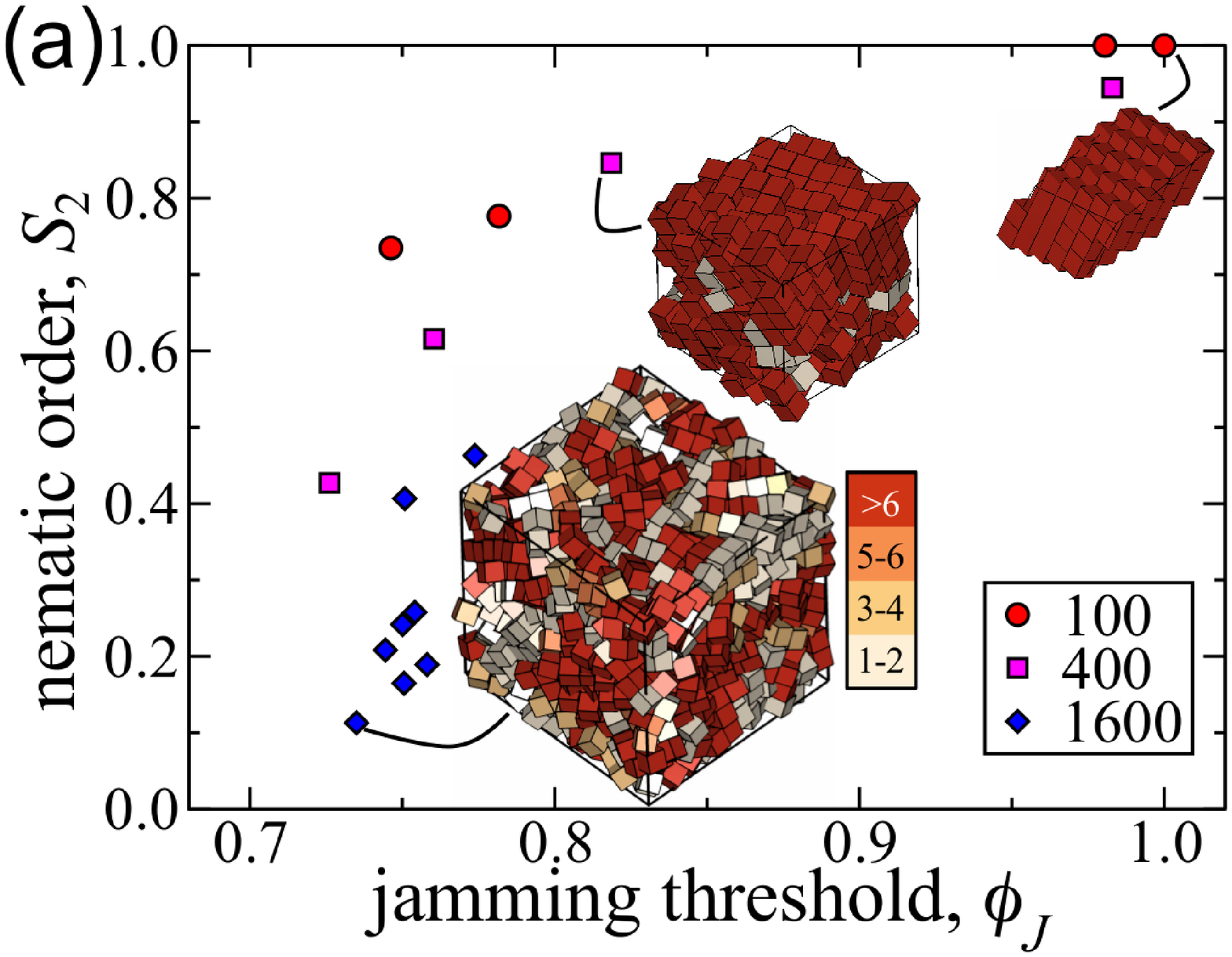}\\
\includegraphics[width=6.5cm]{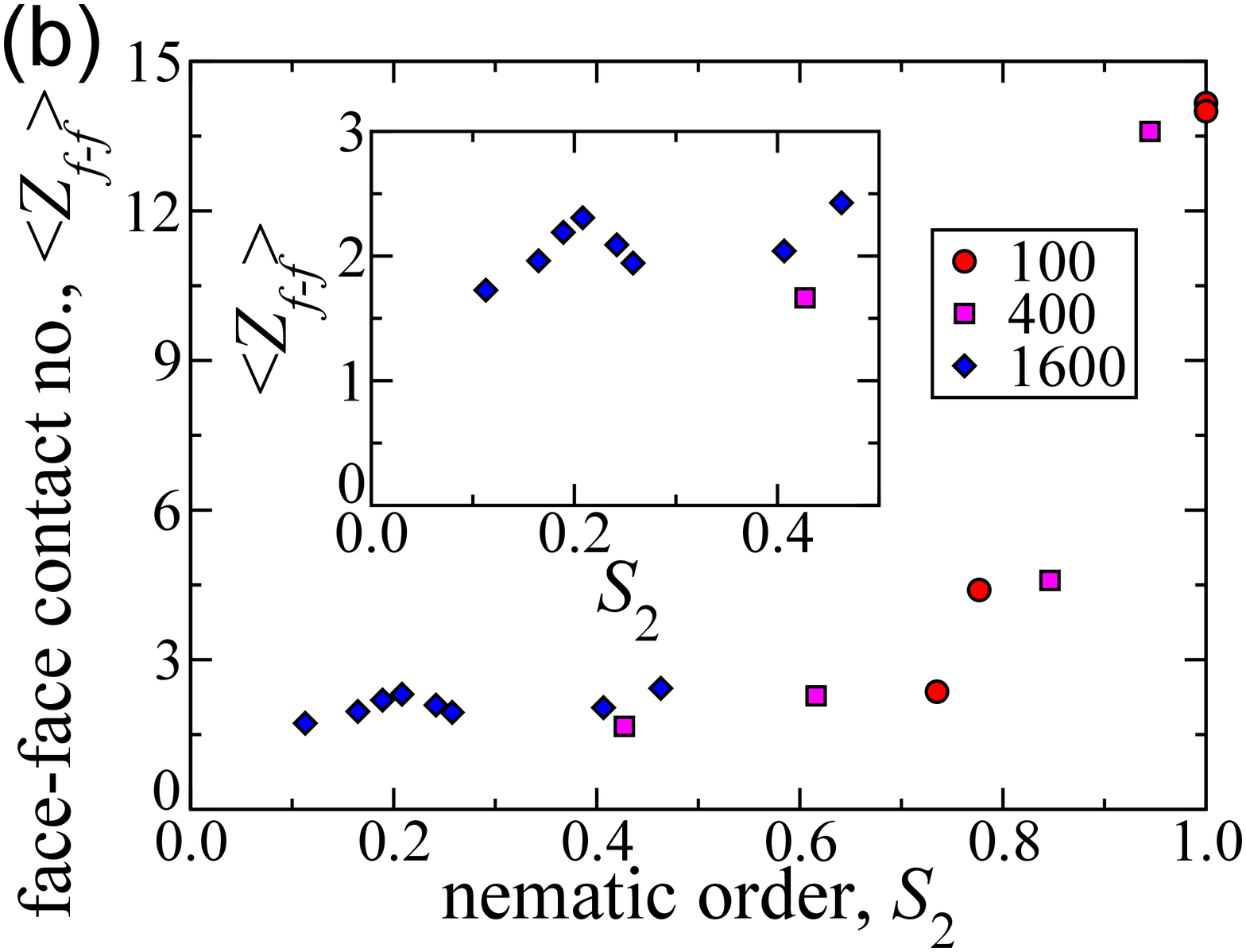}
\caption{(a) Nematic order parameter $S_2$ as a function of jamming threshold density $\phi_J$ for realizations of cubes jammed hydrostatically at $p_{ext}=10^{-4}Y$ in a variable cell.  Thumbnails of selected configurations are depicted with color according to the number of cubes belonging to the face-face cluster to which a given cube corresponds.  Black edges indicate the boundary of the periodic cell. (b) Face-face contact number $\langle Z_{f-f} \rangle$ as a function of nematic order parameter.  The inset shows the variation of $\langle Z_{f-f} \rangle$ for small values of $S_2$.}
\label{fig:cubes}
\end{figure}

\begin{figure}
\centering
\includegraphics[width=6.5cm]{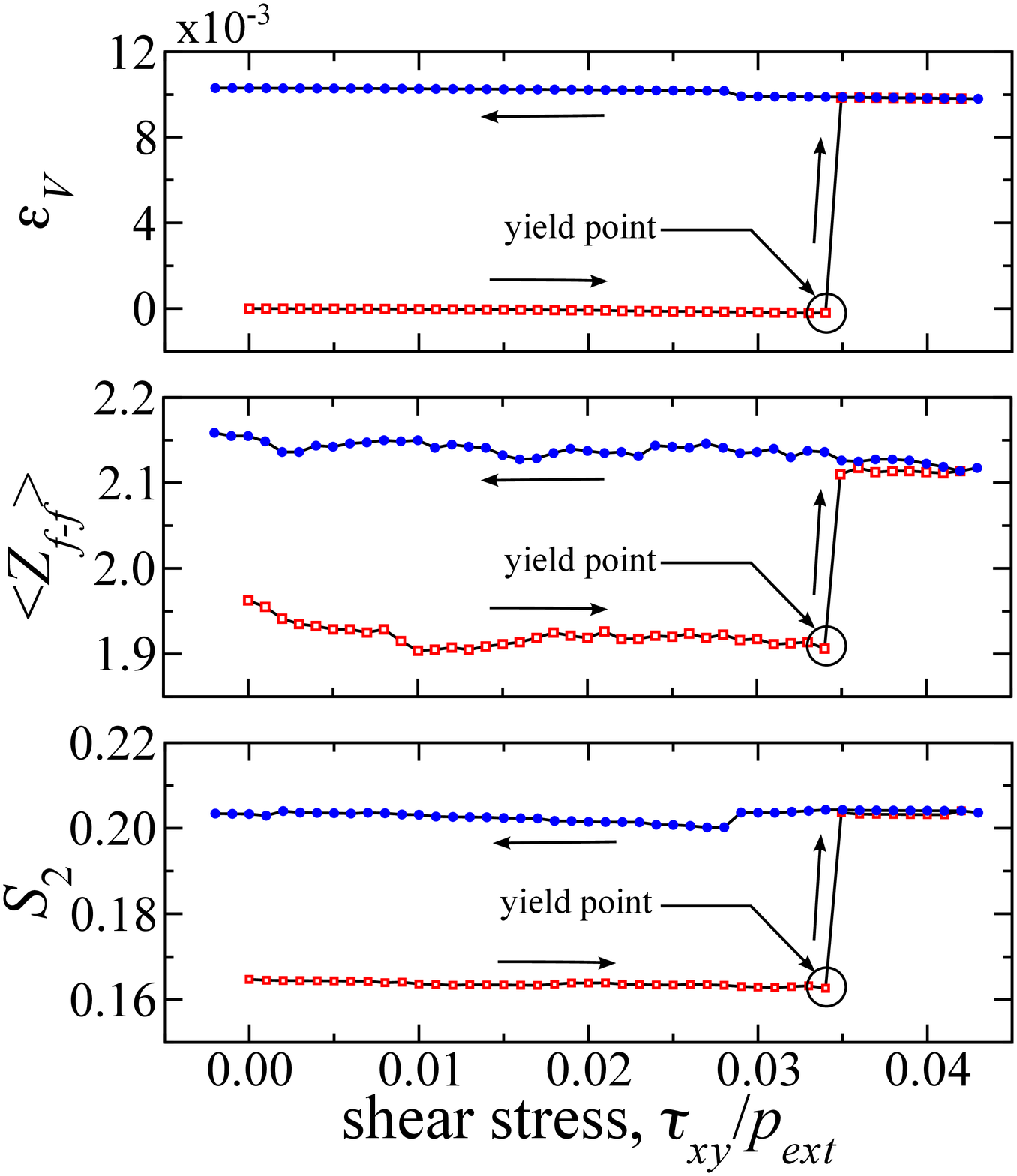}
\caption{Volumetric strain $\epsilon_{V}$, face-face contact number $\langle Z_{f-f} \rangle$, and nematic order $S_2$ as a function of applied shear-stress for a system of $1600$ cubes, at $p_{ext}=10^{-4}Y$.}
\label{fig:cubes2}
\end{figure}

Packed cubes exhibit a variety of phases, including ordered, space-filling phases \cite{TorqJiaoPRE2009}, but jammed phases also form at a threshold density of $\sim0.8$ with strong nematic order in fixed-shape cells of finite size \cite{SmiPRE2010}.  Smooth grain shapes that resemble cubes (e.g., superellipsoids/superballs which interpolate between spheres and cubes) have exhibited similar nematic ordering in cells of fixed shape \cite{DelEPL2010,TorqJiaoPRE2010mrjsuper}.  Cubes exhibit three possible nematic ordering directions aligned with cubic faces, denoted as $\hat{u}$.   By sorting the three orientation directions of a given cube to different sets based on their alignment with the three nematic directors, a nematic tensor can be computed for each director \cite{BetJCP2004}:

\begin{equation}
  \boldsymbol{Q}=N^{-1}\sum_{i=1}^{N}\left(\frac{3}{2}\hat{u}_{i}\otimes\hat{u}_{i}-\frac{1}{2}\boldsymbol{I}_3\right),
\end{equation}

\noindent where $i$ denotes a given grain and the sum loops over all $N$ cubes.  From the dominant eigenvalue of a given nematic tensor $\lambda_{max}$ the uniaxial nematic order parameter is $S_2=2\lambda_{max}-1$.  We report the largest such value among all three nematic directors.

Four realizations of each system size ($100$, $400$, and $1600$ cubes) were simulated.  $100$ cube systems tend to form highly ordered structures as a result of the correlated motions between periodic images.  As a result, two of the four small systems exhibit artificially high density, and the cell shape of these systems strongly deviates from that of a cubic cell [Fig.~\ref{fig:cubes}(a), upper right].  For packings of the most cube-like superellipsoids reported in Ref.~\cite{DelEPL2010} the average uniaxial nematic order parameter was $0.75$; the system of $400$ cubes jammed with a fixed cubic cell in our previous work \cite{SmiPRE2010} exhibited a dominant uniaxial nematic order parameter of $0.74$.  Even more ordered crystalline structures have been simulated in thermalized systems \cite{AgaNMAT2011,SmaPNAS2012}.  In contrast, the realizations of $400$ cubes jammed hydrostatically are spread over a large range of order [see Fig.~\ref{fig:cubes}(a)] that includes the aforementioned values from the literature.  A trend of decreasing order with density is apparent, and six of eight 1600-cube realizations exhibit $S_2\approx0.2$, which reflects dramatically less nematic order than jammed cube-like grains from the previous reports already described.  These cases are indictative of the order expected in the large-system limit.

Despite the lack of long-range nematic order [Fig.~\ref{fig:cubes}(a)], large systems of cubes possess an abundance of face-face contacts [approximately 2 per cube on average assuming that contacts with faces aligned by $<1^{\circ}$ are face-face contacts, Fig.~\ref{fig:cubes}(b)].  These contacts form clusters that can be similar in size to the simulation cell, and a single cluster can contain the majority of grains in small systems [Fig.~\ref{fig:cubes}(a)].  For the largest systems simulated, though, small clusters having 3-4 grains are abundant in these structures.  These clusters may function as steric defects that frustrate the nematic order of jammed cubes.  For sufficiently large $\langle Z_{f-f} \rangle$, clusters can percolate (note that spanning the simulation cell is an insufficient condition for percolation in periodic systems \cite{MakPRE1995}), and these results motivate the exploration of larger systems in which the finite-size scaling of face-face clusters may be quantified.

From a composite-materials engineering perspective, the ability to `tune' cluster percolation with grain shape is useful, because transport processes (e.g., heat, mass, and charge transport) are highly sensitive to the extent of clusters (see Refs.~\onlinecite{SmiPCCP2012,SmiJHT2012,SriJHT2013,SmiIJHE2012}).  One way to control cluster percolation is to shear hydrostatically jammed structures, and thereby increase order, density, and face-face contact number.  To test this hypothesis we have also used the stress-controlled, variable-cell method to simulate the response to large shear-stress perturbations of $1600$ cubes jammed hydrostatically in a variable cell, as shown in Fig.~\ref{fig:cubes2}.  As shear-stress magnitude increases, mild dilation occurs with near-constant $\langle Z_{f-f} \rangle$ and nematic order until the system yields at $\tau_{xy}/p_{ext}\approx0.035$, after which the system contracts, gains more face-face contacts, and becomes more nematic.  As shear-stress magnitude is relaxed toward the initial hydrostatic state-of-stress (i.e., $\tau_{xy}=0$), the system does not return to the initial level of strain and nematic order before it was sheared.  This result suggests that nematic order and face-face contact number in cube systems are senstive to shear stresses.  Thus, shear stress may be used to generate dense, inter-connected assemblies of cubes.

\subsection{Jamming of octahedrally symmetry grains: Face-face contact emergence}\label{sec:cuboct}

\begin{figure}
\centering
\includegraphics[width=6.5cm]{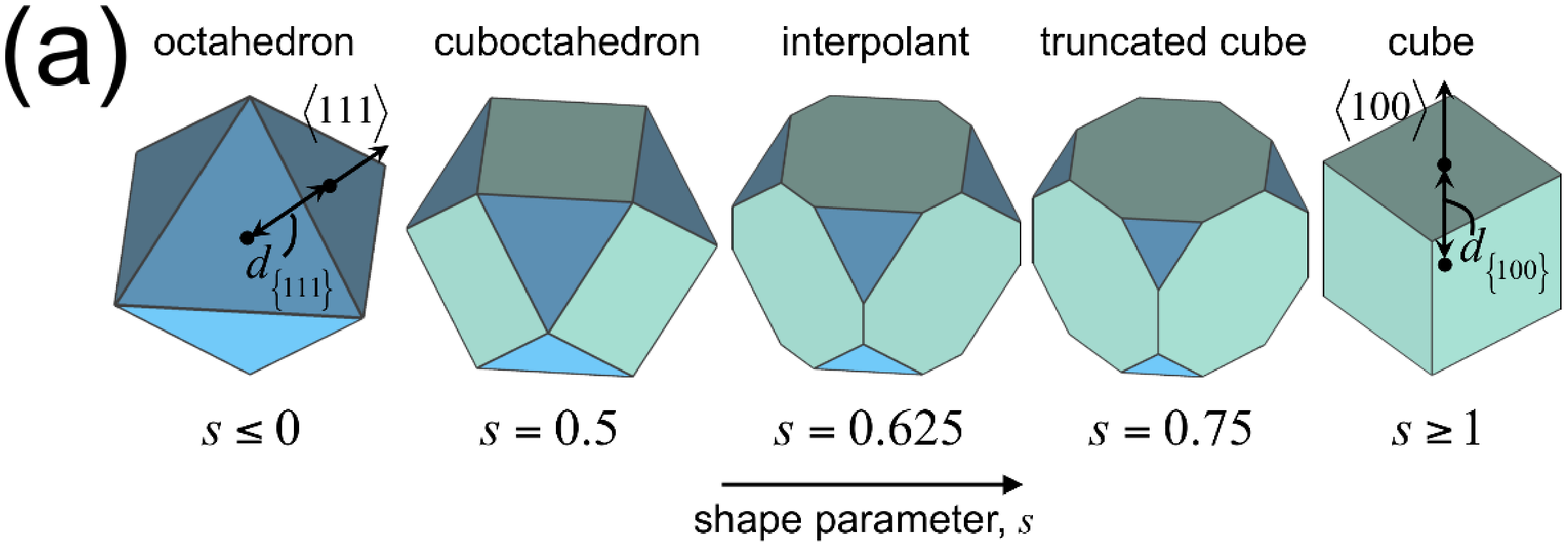}\\
\includegraphics[width=6.5cm]{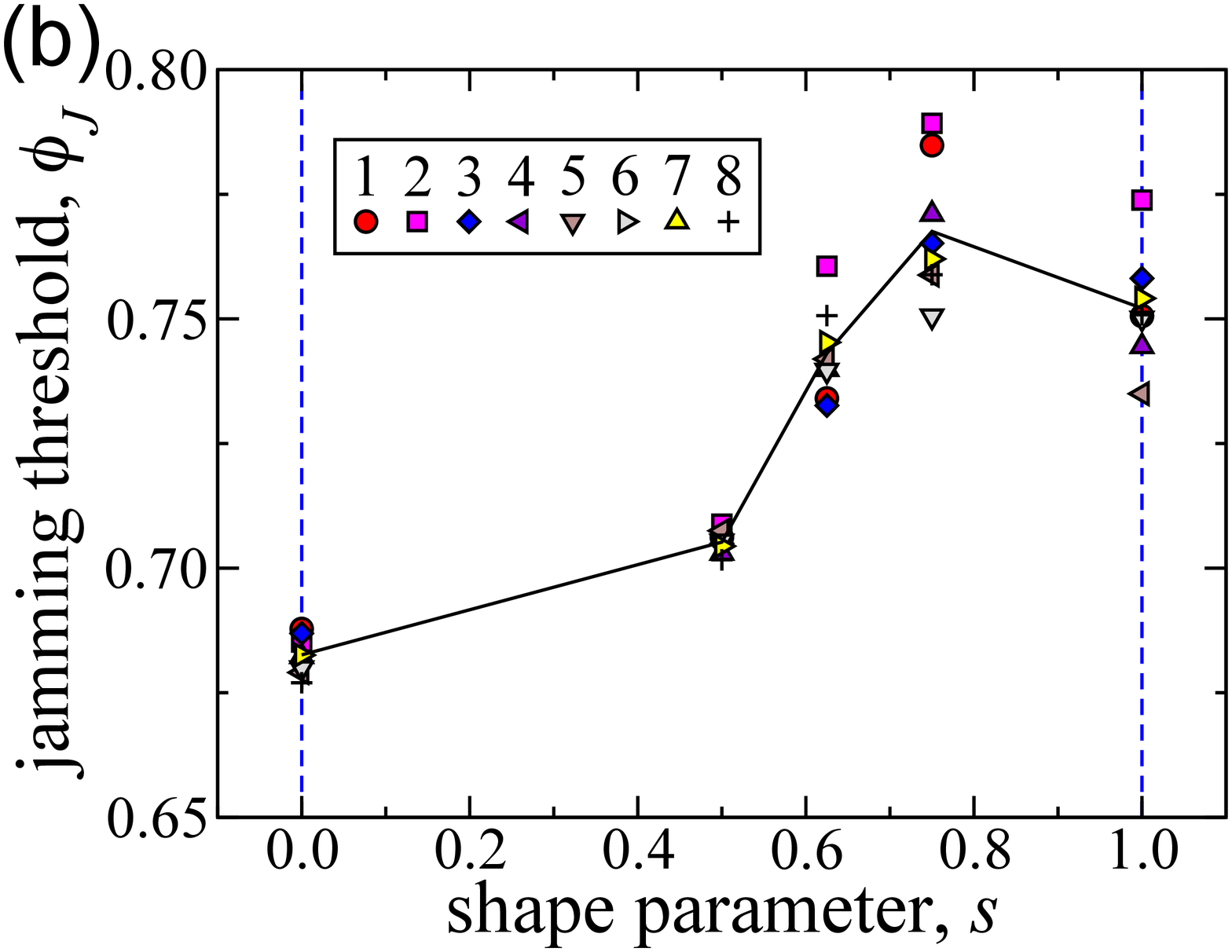}
\caption{(a) Octahedrally symmetric grain-shape as a function of shape parameter $s$.  Green and blue faces lie on $\{100\}$ and $\{111\}$ crystallographic planes, respectively, and the directions normal to those planes are indicated.  (b) Jamming threshold density $\phi_J$ as a function of shape parameter $s$ for various realizations of $1600$ grains (specified by seed number indicated in the legend).  Black lines were averaged over all realizations of a given grain shape.}
\label{fig:cuboct_density}
\end{figure}

\begin{figure}
\centering
\includegraphics[width=6.5cm]{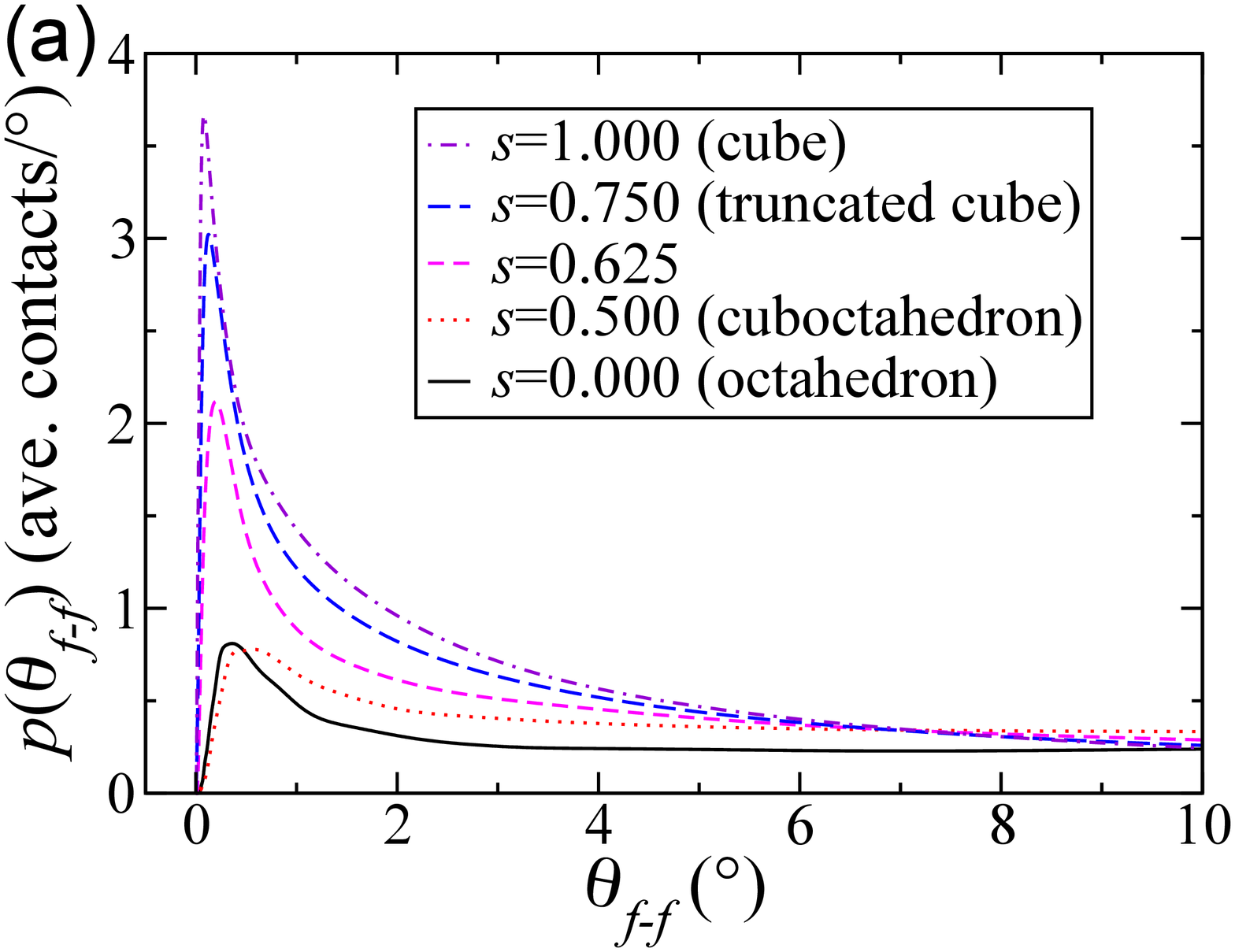}\\
\includegraphics[width=6.5cm]{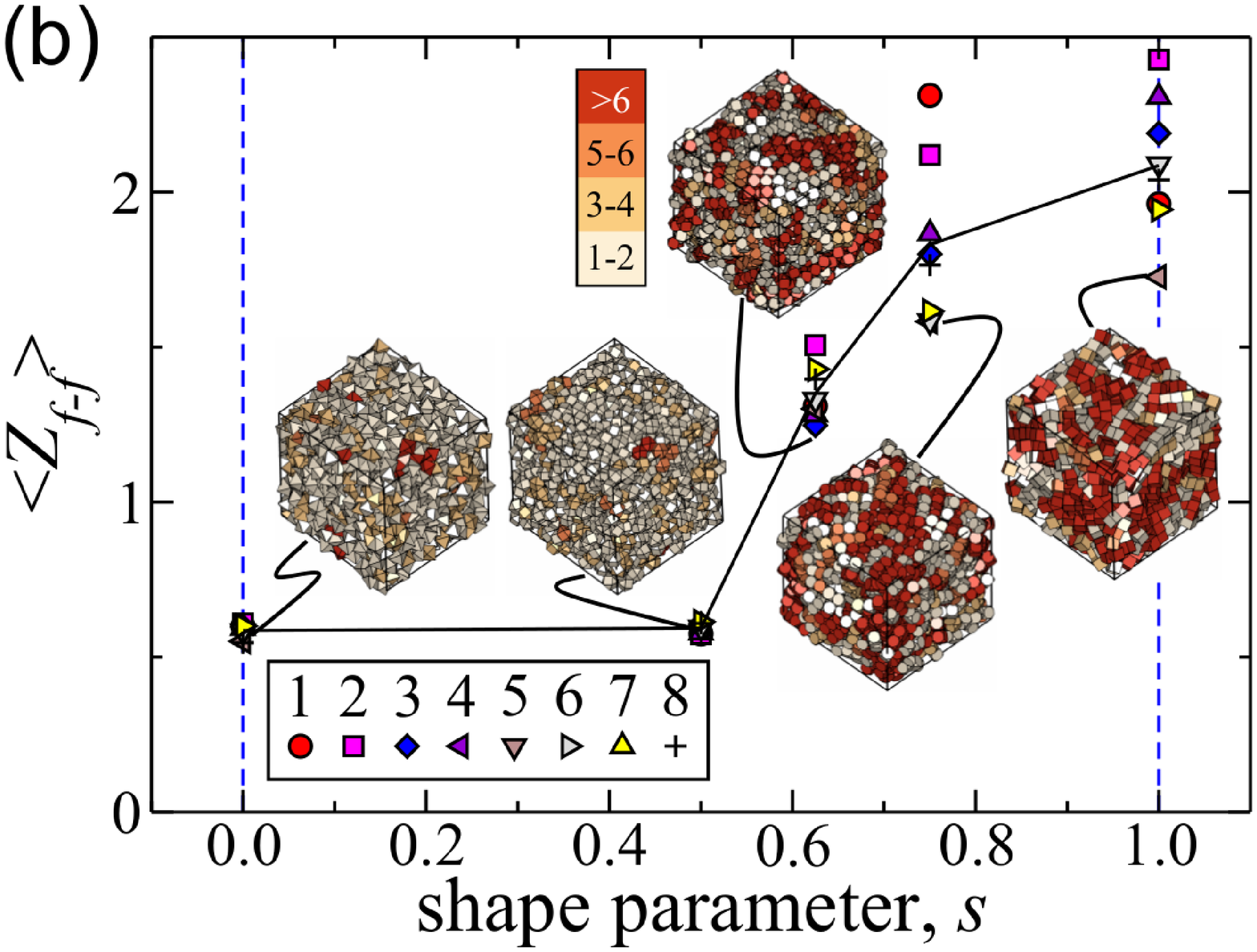}\\
\includegraphics[width=6.5cm]{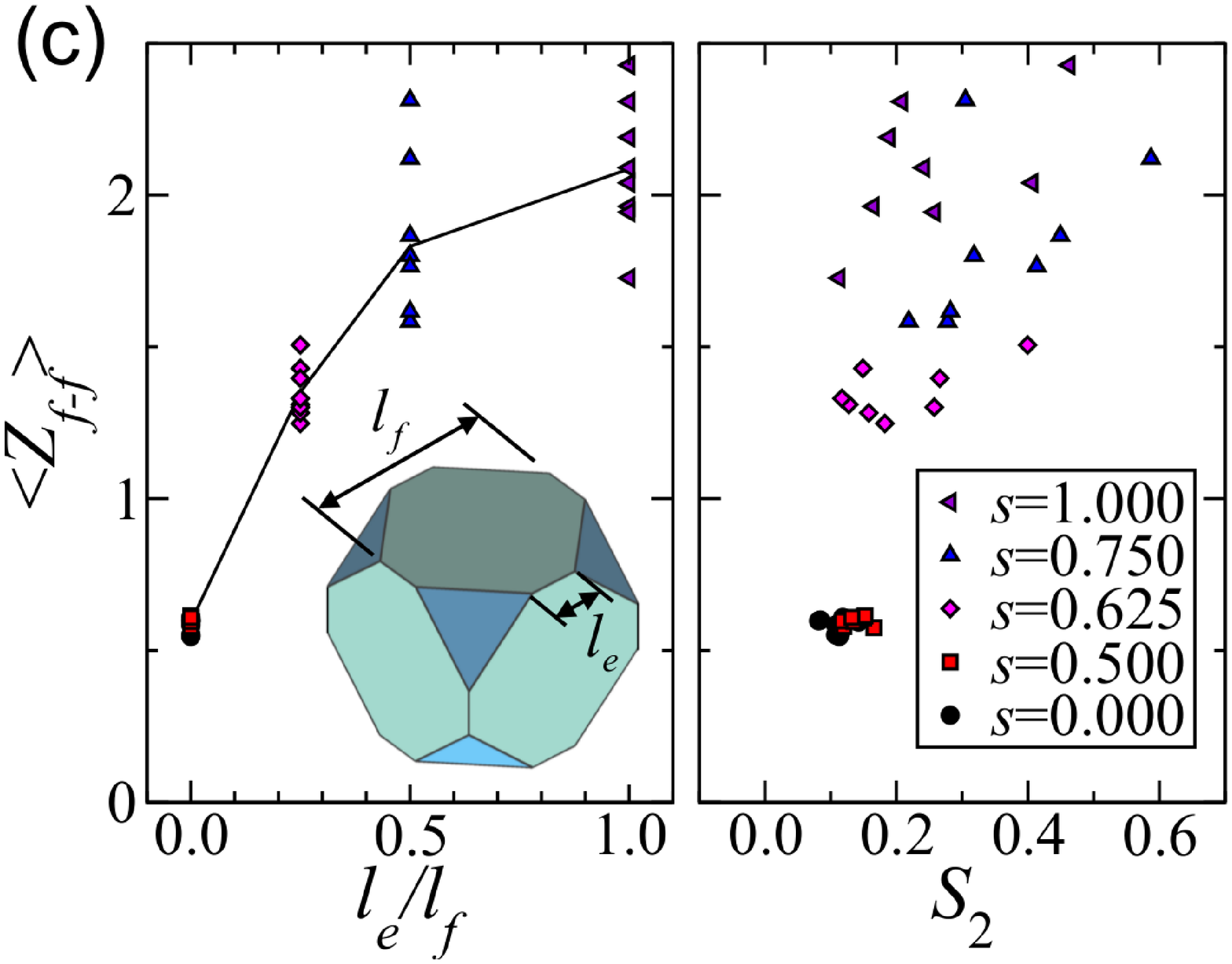}
\caption{(a) Probability distribution function of face-face alignment angle for jammed systems of $1600$ octahedrally symmetric grains at $p_{ext}=10^{-4}Y$.  These functions were averaged over four realizations and were computed by bounded kernel density estimation \cite{bndkde}.  (b) Average face-face contact number $\langle Z_{f-f} \rangle$ for each realization (specified by seed number indicated in the legend) of a given shape.  Grains are colored (see legend) according to the size of the face-face clusters to which they belong for the selected realization (black lines indicate periodic cell boundaries).  (c) Average face-face contact number as a function of edge-to-face length-ratio $l_e/l_f$.  Black lines were averaged over all realizations of a given grain shape.}
\label{fig:cuboct}
\end{figure}

Because of their dual symmetry, cubes and octahedra belong to a class of grain shapes parameterized by a linear interpolation variable $s$:

\begin{equation}
  s=0.5(\sqrt{3}d_{\{111\}}/d_{\{100\}}-1),
\end{equation}

\noindent where $d_{\{111\}}$ and $d_{\{100\}}$ are the distances from the grain's centroid to $\{111\}$ and $\{100\}$ crystallographic faces of a given grain shape.   As shown in Fig.~\ref{fig:cuboct_density}(a), the octahedron ($s=0$) and cube ($s=1$) are extremes in this class and several of the Archimedean solids interpolate between them: the truncated octahedron ($s=0.25$, not shown), cuboctahedron ($s=0.5$), and truncated cube ($s=0.75$).  Both the Platonic and Archimedean solids have recently received attention with regard to optimal \cite{TorNat2009} and jammed \cite{SmiPRE2010,SmiPRE2011,BakKudPRE2010} packing.  Also, the self-assembly of thermalized grains belonging to this class have been explored recently \cite{AgaJCP2012,VolNAN2012}.  Knowledge of how these grains jam at zero temperature will yield insight into the kinetic barriers that inhibit ordering in thermalized systems.  Also, this class of grains is practically useful for bottom-up composite material fabrication, because crystallographic structures having octahedral symmetry (e.g., simple cubic or diamond) can form faceted grain shapes with the same symmetries.

Eight realizations of $1600$ grains of various shapes having octahedral symmetry are simulated under a hydrostatic pressure of $p_{ext}=10^{-4}Y$ in a variable cell.  The jamming threshold density $\phi_J$ exhibits a marked increase as $s$ increases near $s\approx0.5$, see Fig.~\ref{fig:cuboct_density}(b).  We also estimated face-face contact numbers $\langle Z_{f-f} \rangle$ (i.e., average number of face-face contacts per grain) based upon the statistics of face-face alignment angles, as in Ref.~\onlinecite{SmiPRE2011}.  While jammed cubes exhibit an abundance of face-face contacts, our previous results suggest that octahedra jam with very few face-face contacts ($\sim 1$ per grain on average in a cube-shaped cell \cite{SmiPRE2011}).  Among the various octahedrally symmetric shapes we have considered, a strong peak in the probability distribution of face-face alignment angle $\theta_{f-f}$ (see Ref.~\onlinecite{SmiPRE2011}) emerges for shapes with $s>0.5$, as in Fig.~\ref{fig:cuboct}(a).  Assuming that contacts with $\theta_{f-f}<1^{\circ}$ are face-face contacts, we estimate the average face-face contact number, as shown in Fig.~\ref{fig:cuboct}(b).  For $0 \le s \le 0.5$ the average face-face contact number $\langle Z_{f-f} \rangle$ is constant at $\approx 0.6$, but for shapes with $s>0.5$, $\langle Z_{f-f} \rangle$ increases linearly and saturates to $\approx2$ contacts per grain.  Recall that the jamming threshold density $\phi_J$ [Fig.~\ref{fig:cuboct_density}(b)] follows a similar trend with $s$, suggesting a correlation between $\langle Z_{f-f} \rangle$ and $\phi_J$.

The emergence of many face-face contacts has a dramatic effect on microstructure.  In particular, the size of clusters formed by grains connected via face-face contacts has a strong dependence on the number of face-face contacts.  The extent of these clusters can have a strong impact on the shear response \cite{SmiPRE2011} and the transmission of heat/charge through the granular medium \cite{SmiJHT2012,SriJHT2013}.  For $s\leq0.5$ the microstructures are comprised of minimal-length clusters (dimers having less than two grains) [Fig.~\ref{fig:cuboct}(b)].  For $s>0.5$ the microstructures are comprised of a large variety of cluster lengths [Fig.~\ref{fig:cuboct}(b)], which are long enough in some cases to span the entire simulation cell.  This result contrasts strongly with the structures observed for other Platonic solids that did not exhibit face-face cluster percolation in our previous fixed-cell studies \cite{SmiPRE2011,SmiJHT2012}.

Though ordering can induce face-face contact formation, many face-face contacts can form in systems with negligible nematic order (i.e., for $s>0.5$).  A scatter plot of $\langle Z_{f-f} \rangle$ versus $S_2$ [right panel of Fig.~\ref{fig:cuboct}(c)] shows that these two parameters are uncorrelated and are not causally related.  Even with $S_2<0.2$, substantial $\langle Z_{f-f} \rangle$ values are observed.  In contrast, a correlated trend of increasing $\langle Z_{f-f} \rangle$ with the length of edges shared by $\{100\}$ faces on each shape is observed [left panel of Fig.~\ref{fig:cuboct}(c)].  Only for $s>0.5$ does such an edge, which is aligned with a $\langle100\rangle$ direction, appear in the shape's topology.  This correlation suggests that the presence of such $\langle100\rangle$ edges is necessary to form face-face contacts in excess of one per grain (on average).

\section{Conclusions}

A general variable-cell method for the stress-based jamming of soft, frictionless grains has been introduced via an $NPH$-type ensemble.  The hydrostatic jamming process simulated at zero temperature allows for the probing of the ideal \textit{jamming point} at zero shear stress specified precisely on the granular phase-diagram, and yields structures that are highly disordered in the large-system limit.  Specifically, jamming under hydrostatic conditions produces structures of tetrahedra with less translational order and cubes with less nematic order than observed previously.  The structures of grains with octahedrally symmetric shapes formed by hydrostatic jamming exhibit face-face contacts as large as triple the values exhibited by other shapes (e.g., tetrahedra \cite{SmiPRE2011}) promoted by the presence of edges shared among intersecting $\{100\}$ faces, that can form percolating clusters which span the whole simulation box in systems.  In addition, the versatility of the present method to simulate jamming packings under both hydrostatic and shear loadings has been demonstrated.

\section{Acknowledgements}

The authors acknowledge the Indo-U.S. Science and Technology Forum for supporting Purdue-JNCASR exchanges through the Joint Networked Centre on Nanomaterials for Energy (Award No. 115-2008/2009-10).


\bibliography{mainbib} 

\end{document}